\documentclass[twocolumn,english,twocolumn,english,aps,prl,floats]{revtex4}
\usepackage[T1]{fontenc}
\usepackage[latin9]{inputenc}
\usepackage{bm}
\usepackage{amsmath}
\usepackage{amssymb}
\usepackage{graphicx}
\usepackage{esint}

\makeatletter
\@ifundefined{textcolor}{}
{%
 \definecolor{BLACK}{gray}{0}
 \definecolor{WHITE}{gray}{1}
 \definecolor{RED}{rgb}{1,0,0}
 \definecolor{GREEN}{rgb}{0,1,0}
 \definecolor{BLUE}{rgb}{0,0,1}
 \definecolor{CYAN}{cmyk}{1,0,0,0}
 \definecolor{MAGENTA}{cmyk}{0,1,0,0}
 \definecolor{YELLOW}{cmyk}{0,0,1,0}
}

\usepackage{babel}

\makeatother

\usepackage{babel}
\begin{document}

\title{Conservation of Angular Momentum in the Elastic Medium with Spins}

\author{Dmitry A. Garanin and Eugene M. Chudnovsky}

\affiliation{Physics Department, Herbert H. Lehman College and Graduate School,
The City University of New York, 250 Bedford Park Boulevard West,
Bronx, New York 10468-1589, USA }

\date{\today}
\begin{abstract}
Exact conservation of the angular momentum is worked out for an elastic medium with spins. The intrinsic anharmonicity of the elastic theory is shown to be crucial for conserving the total momentum. As a result, any spin-lattice dynamics inevitably involves multiphonon processes and interaction between phonons. This makes transitions between spin states in a solid fundamentally different from transitions between atomic states in vacuum governed by linear electrodynamics. Consequences for using solid-state spins as qubits are discussed.
\end{abstract}

\maketitle

The problem of transfer of angular momentum between magnetic moments and macroscopic body goes back to seminal experiments of Einstein --
de Haas \cite{EdH} and Barnett \cite{Barnett}. The first established that the change in the magnetization of a freely suspended body is accompanied by mechanical rotation. The second demonstrated that rotation causes magnetization. Macroscopic explanation of these phenomena is straightforward -- based upon conservation of the total angular momentum. Equally straightforward is microscopic theory of spin-phonon processes developed in seminal papers of Van Vleck \cite{VanVleck-PR40}. Magnetoelastic Hamiltonians studied by classical and quantum methods for specific materials ever since have been written to reflect anisotropy of the crystal lattice. Due to the lack of rotational symmetry they do not conserve the total angular momentum. Until recently this inconsistency was swept under the rug by correctly assuming that any unaccounted change in the angular momentum would be absorbed by the whole body. Attempts to demonstrate this in a rigorous manner \cite{Melcher,Dohm,Fedders}, while conceptually valuable, have been mathematically cumbersome with no clear consequences for experiments. 

In recent years this problem received renewed attention due to the emergence of spintronics and nanoelectromechanical devices, as well as due to the prospect of developing spin-based quantum computers. Universal parameter-independent nature of spin-lattice relaxation arising from the conservation of the total angular momentum has been demonstrated \cite{chugarsch05prb}. Transfer of the angular momentum from spins to mechanical degrees of freedom in nanomechanical oscillators has been studied theoretically \cite{Kovalev-PRL2005,jaachugar09prb,jaachu09prl,kovhaybau11prl,carchu11prx,chugar14prb}  and experimentally \cite{walmorkab06apl,ganklyrub13NatNano,Freeman-2018,Freeman-PRB2020}. The division of the phonon angular momentum into orbital and spin parts was suggested \cite{Zhang-Niu} and further explored in application to problems of spin relaxation \cite{GC-PRB2015,Nakane2018} and spin transport \cite{Duine2020}. The concept of angulon, a quasiparticle carrying an angular momentum, initially introduced to describe properties of molecular impurities in a superfluid \cite{Lemeshko2017}, has been extended to magnetic impurities in solids \cite{Katsnelson2019}. The physics of the Einstein - de Haas effect in magnetic insulators has been recently revisited within a model that decouples rotations from vibrations and separates variables responsible for microscopic and macroscopic mechanical torques \cite{Bauer2020}.

In Ref.\ \onlinecite{GC-PRB2015} we demonstrated conservation of the angular momentum by spin-lattice interaction for a specific quantum problem of a relaxing spin. The general case turned out to be more subtle even at the classical level. It will be solved here and will elucidate the fact that spin-phonon processes are fundamentally different from spin-photon processes due to the intrinsic nonlinearity of the elastic theory as compared to electrodynamics. 

We begin with underappreciated derivation of the conservation of angular momentum in a conventional elastic theory that is not easy to find in textbooks. The expression for the angular momentum of the elastic solid that is linear on a small deformation $\mathbf{u}({\bf r},t)$ is:
\begin{equation}
\mathbf{L}^{(0)}=\int d^{3}r\left(\mathbf{r}\times\mathbf{p}\right).\label{L0}
\end{equation}
Here $\mathbf{p}({\bf r},t)=\rho\dot{{\bf u}}({\bf r},t)$ is the momentum density, with $\rho$ being the mass density of the solid. The time derivative of this expression yeilds:
\begin{equation}
\dot{\bf L}^{(0)} = \int d^3 r \rho ({\bf r}   \times \ddot {\bf u}). \label{L0-derivative}
\end{equation}
The second time derivative of ${\bf u}$ in this equation satisfies the Newton's equation,
\begin{equation}
\rho \frac{\partial^2 u_{\alpha}}{\partial t^2} = \frac{\partial \sigma_{\alpha \beta}}{\partial r_{\beta}}, \label{Newton}
\end{equation}
with the force in the right-hand-side being the gradient of the stress tensor  $\sigma_{\alpha \beta} = {\delta {\cal{H}}}/{\delta e_{\alpha\beta}}$. Here ${\cal{H}}$ is the Hamiltonian of the system and $e_{\alpha \beta} = {\partial u_{\alpha}}/{\partial r_{\beta}}$ is the strain tensor. Substituting Eq.\ (\ref{Newton}) into Eq.\ (\ref{L0-derivative}) and integrating by parts under the assumption of zero elastic stress at the free boundary of the body, we obtain
\begin{equation}
\dot{L}^{(0)}_{\alpha} = \int d^3 r \epsilon_{\alpha \beta \gamma} r_{\beta} \frac{\partial
\sigma_{\gamma \delta}}{\partial r_{\delta}} =  -\int d^3 r  \epsilon_{\alpha \beta \gamma}\sigma_{\gamma\beta}, \label{byparts}
\end{equation}
where summation over repeating indices is assumed with $\epsilon_{\alpha \beta \gamma}$ being an absolutely antisymmetric unit tensor of third rank (Levi-Civita symbol). In the linear theory of elasticity that ignores local internal torques associated with spins the Hamiltonian is 
\begin{equation}
{\cal{H}} = \int d^3 r \left(\frac{1}{2}\rho \dot{{\bf u}}^2 +  \frac{1}{2}C_{\alpha\beta\gamma\delta}u_{\alpha\beta}u_{\gamma\delta}\right) \label{H}
\end{equation}
with $u_{\alpha\beta} = \frac{1}{2}(e_{\alpha\beta} + e_{\beta\alpha})$ and tensor $C_{\alpha\beta\gamma\delta}$ reflecting the symmetry of the crystal lattice. This makes the stress tensor $\sigma_{\gamma\beta}$  in Eq.\ (\ref{byparts}) proportional to the strain (Hooke's law) and symmetric with respect to the transposition of $\gamma$ and $\beta$. The product of the antisymmetric tensor $\epsilon_{\alpha \beta \gamma}$ and the symmetric tensor $\sigma_{\gamma\beta}$ in the last of Eq.\ (\ref{byparts}) is automatically zero, which makes $\dot{\bf L}^{(0)}$ zero and proves conservation of the angular momentum in the limit linear on deformations. 

This however is not the end of the story if one recalls the exact definition of the symmetrized strain tensor,
\begin{equation}
u_{\alpha\beta} = \frac{1}{2}\left(e_{\alpha\beta} + e_{\beta\alpha} + e_{\gamma\alpha}e_{\gamma\beta}\right), \label{strain-nonlinear}
\end{equation} 
that follows from the analysis  \cite{LL-elasticity} of how deformations change distances between two points in the elastic medium:
\begin{equation}
dl'^{2}=dl^{2}+2u_{\alpha\beta}dr_{\alpha}dr_{\beta}. \label{length}
\end{equation}
With that definition of $u_{\alpha\beta}$ one obtains
\begin{equation}
\sigma_{\alpha\beta} = \frac{\delta {\cal{H}}}{\delta e_{\alpha\beta}} = \frac{\delta {\cal{H}}}{\delta u_{\alpha\beta}} + e_{\alpha\gamma}\frac{\delta {\cal{H}}}{\delta u_{\gamma\beta}}. \label{full-sigma}
\end{equation}
The first tensor in this expression is symmetric but the second is not, even for an isotropic body, which makes  $\dot{\bf L}_0$ in Eq.\ (\ref{byparts}) nonzero. Solution to this problem comes from the realization that the full expression for the angular momentum in the elastic medium is
\begin{equation}
\mathbf{L}=\int d^{3}r\left(\mathbf{r}+\mathbf{u}\right)\times\mathbf{p} = \int d^3 r \rho ({\bf r}  + {\bf u})\times \dot {\bf u}.\label{full-L}
\end{equation}
The term quadratic on ${\bf u}$, that Zhang and Niu \cite{Zhang-Niu} associated with the phonon spin, comes from the same necessity to distinguish between coordinates of atoms before and after deformation that leads to Eq.\ (\ref{strain-nonlinear}) via Eq.\ (\ref{length}). The time derivative of the angular momentum now becomes
\begin{equation}
\dot{L}_{\alpha} = -\int d^3 r \left( \epsilon_{\alpha \beta \gamma}\sigma_{\gamma\beta} +\epsilon_{\alpha \beta \gamma} e_{\beta\delta}\sigma_{\gamma\delta} \right). \label{dotL-sigma}
\end{equation} 
Substituting Eq.\ (\ref{full-sigma}) into Eq.\ (\ref{dotL-sigma}) and working out tensor products that contain zero convolutions of symmetric and antisymmetric tensors, one obtains $\dot{\bf L} = 0$, which proves the exact conservation of the angular momentum in the conventional nonlinear elastic theory to all orders on the deformation. 

\begin{figure}[h]
\centering{}
\includegraphics[width=13.0cm]{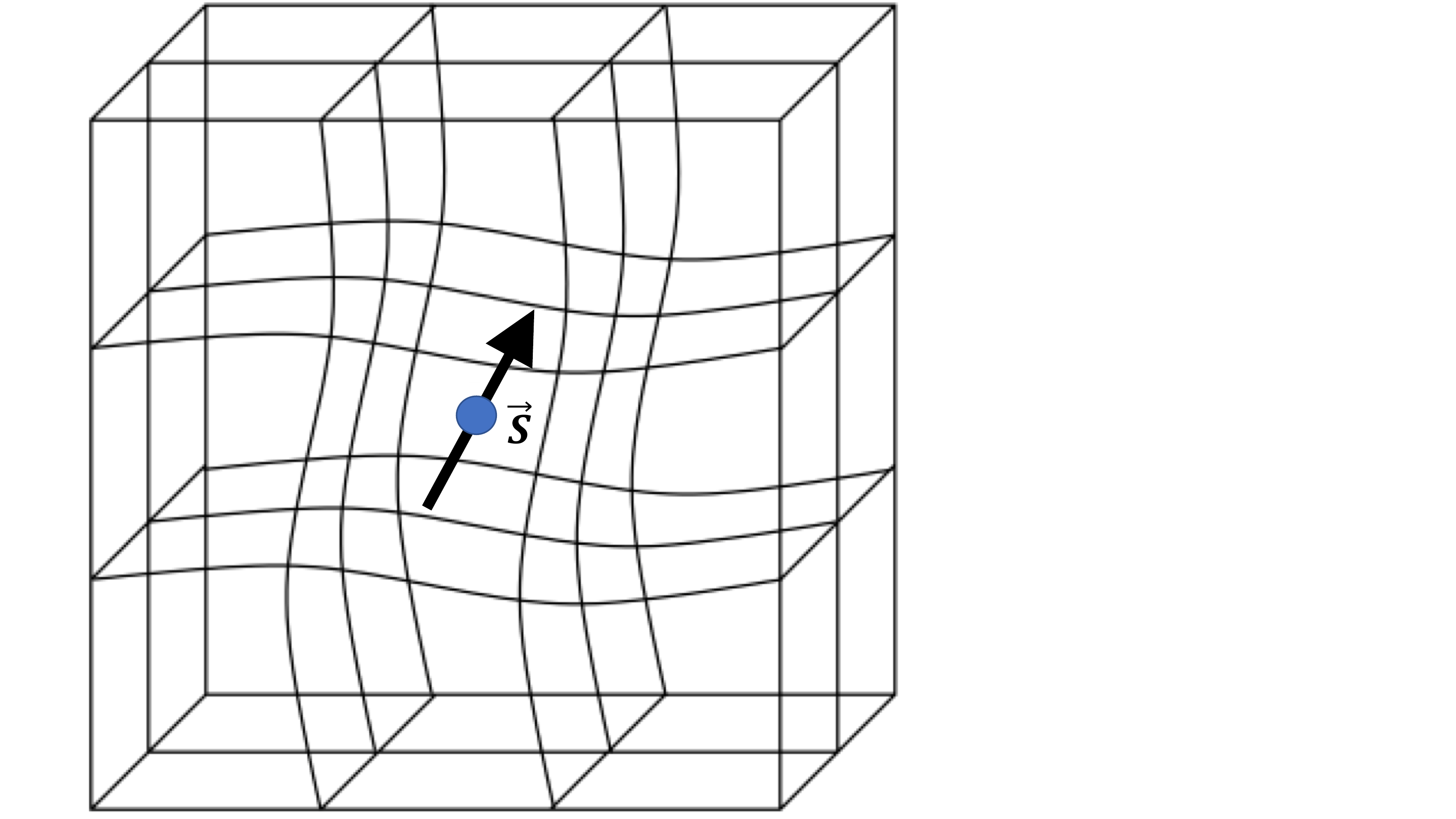} 
\caption{Schematic illustration of the elastic twist due to the reaction of the atomic lattice to a rotating spin.}
\label{fig1}
\end{figure}
We shall now include into the theory a single atomic spin located at a point ${\bf r} = {\bf r}_0$. This immediately takes us outside the framework of the conventional elasticity because spins, as they rotate, generate internal torques (see Fig.\ \ref{fig1}) that are explicitly neglected by the elastic theory. As we shall see, however, this problem can be fixed in the same manner as the linear theory of elasticity was fixed before. 

Consider for certainty a uniaxial spin Hamiltonian 
\begin{equation}
{\cal{H}}_{S} = {H}_{S}\delta(\mathbf{r}-\mathbf{r}_0), \quad {H}_{S} = F\left(\mathbf{n\cdot S}\right),  \label{Ham_S}
\end{equation}
where $F$ is an arbitrary scalar function of its argument, e.g., $F=-D\left(\mathbf{n\cdot S}\right)^{2}$,
with ${\bf n}$ being the magnetic anisotropy axis in the non-deformed
state. Elastic deformations of the body
change the direction of the anisotropy axis $\mathbf{n}$. Local rotation by a small angle $\delta {\bm \phi} = \frac{1}{2} {\bm \nabla} \times {\bf u}$ changes $\mathbf{n}$ to
\begin{equation}
{\bf n}' = {\bf n} +{\delta {\bm \phi}}\times{\bf n} + \frac{1}{2}{\delta{\bm \phi}}\times({\delta{\bm \phi}}\times{\bf n}) + ...
\end{equation}
In the first order on deformation it results in the spin-lattice interaction of the form
\begin{equation}
\delta H_S = \frac{\partial F}{\partial {\bf n}}\cdot ({\delta {\bm \phi}}\times{\bf n}). \label{HS-linear}
\end{equation}
It is difficult, however, to develop rigorous approach in all orders on $\delta {\bm \phi}$ needed to prove exact conservation of the angular momentum. 

Below we adopt a different approach by noticing that the unit vector ${\bf n}$ determined by the symmetry of the crystal lattice goes in the direction of the vector connecting two points in a solid. Deformation changes it as
\begin{equation}
\mathbf{n}=\frac{d\mathbf{r}}{dl}\Rightarrow\mathbf{n}'=\frac{d\mathbf{r}'}{dl'},
\end{equation}
where
\begin{equation}
dr'_{\alpha}=dr_{\alpha}+du_{\alpha}=dr_{\alpha}+e_{\alpha\beta}dr_{\beta}
\end{equation}
and $dl=|d\mathbf{r}|$ and $dl'=|d\mathbf{r}'|$ are the lengths
of the infinitesimal vector $d\mathbf{r}$ before and after deformation,
related through Eq.\ (\ref{length}). From these formulas one obtains
\begin{equation}
n'_{\alpha}=\frac{d{r}'_{\alpha}}{dl'}= \frac{n_{\alpha}+e_{\alpha\beta}n_{\beta}}{\sqrt{1+2u_{\delta\eta}n_{\delta}n_{\eta}}}.\label{npr_alpha_res} 
\end{equation}
With the replacement of ${\bf n}$ by ${\bf n}'$ the Hamiltonian in Eq.\ (\ref{Ham_S}) becomes the exact spin-lattice Hamiltonian accounting for all orders on the deformation,
\begin{equation}
{H}_{S} = F\left(\mathbf{n'\cdot S}\right),  \label{Exact-Ham_S}
\end{equation}
where ${\bf n}'$ is given by Eq.\ (\ref{npr_alpha_res}).

The spin contribution to the mechanical stress tensor is 
\begin{equation}
\sigma^{(S)}_{\alpha\beta}=\frac{\delta\mathcal{H}_{S}}{\delta e_{\alpha\beta}}=\frac{\delta\mathcal{H}_{S}}{\delta n'_{\gamma}}\frac{\partial n'_{\gamma}}{\partial e_{\alpha\beta}}. \label{sigma-S}
\end{equation}
After a straightforward algebra we get with the help of Eq. (\ref{strain-nonlinear})
\begin{equation}
 \frac{\partial n'_{\alpha}}{\partial e_{\beta\gamma}} =  \frac{(\delta_{\alpha\beta}-n'_{\alpha}n'_{\beta})n_{\gamma}}{{\sqrt{1+2u_{\delta\eta}n_{\delta}n_{\eta}}}}. \label{npr_der_rewritten}
\end{equation}
Substituting this into Eq.\ (\ref{sigma-S}) one obtains from Eq.\ (\ref{dotL-sigma}) the spin contribution to the time derivative of the mechanical angular momentum:
\begin{equation}
\dot{L}^{(S)}_{\alpha} = -\epsilon_{\alpha\beta\gamma}\left(\delta_{\beta\delta}+e_{\beta\delta}\right)\frac{\partial\hat{H}_{S}}{\partial n'_{\eta}}\frac{\partial n'_{\eta}}{\partial e_{\gamma\delta}}. \label{LSdot}
\end{equation}
With the help of Eq.\ (\ref{Ham_S}), it can be re-written as
\begin{equation}
\dot{L}^{(S)}_{\alpha}=-F'\left(\mathbf{\mathbf{n}}'\cdot\mathbf{S}\right)S_{\eta}\epsilon_{\alpha\beta\gamma}\left(\delta_{\beta\delta}+
e_{\beta\delta}\right)\frac{\partial n'_{\eta}}{\partial e_{\gamma\delta}}, \label{L_dot_alpha_res}
\end{equation}
where $ F' (x)=dF/dx$. Eqs.\ (\ref{npr_alpha_res}) and (\ref{npr_der_rewritten}) allow one to write
\begin{equation}
\left(\delta_{\beta\delta}+e_{\beta\delta}\right)\frac{\partial n'_{\eta}}{\partial e_{\gamma\delta}}=n'_{\beta}\left(\delta_{\gamma\eta}-n'_{\eta}n'_{\gamma}\right).
\end{equation}
Dropping the zero convolution of symmetric and antisymmetric tensors we obtain from Eq (\ref{L_dot_alpha_res})
\begin{eqnarray}
\dot{L}^{(S)}_{\alpha} & = & -F'\left(\mathbf{\mathbf{n}}'\cdot\mathbf{S}\right)S_{\eta}\epsilon_{\alpha\beta\gamma}n'_{\beta}\left(\delta_{\gamma\eta}-n'_{\eta}n'_{\gamma}\right) \nonumber \\
& = &  -F'\left(\mathbf{\mathbf{n}}'\cdot\mathbf{S}\right)\epsilon_{\alpha\beta\gamma}n'_{\beta} S_{\gamma}. \label{LS-final}
\end{eqnarray}

The equation of motion for the spin is
\begin{equation}
\hbar\mathbf{\dot{S}}=-\left[\mathbf{S}\times\frac{\delta\mathcal{H}_{S}}{\delta\mathbf{S}}\right],
\end{equation}
or, in the tensor form
\begin{equation}
\hbar\dot{S}_{\alpha}=-\epsilon_{\alpha\beta\gamma}S_{\beta}\frac{\delta\mathcal{H}_{S}}{\delta S_{\gamma}}.
\end{equation}
Using Eq. (\ref{Ham_S}), one can write it as
\begin{equation}
\hbar\dot{S}_{\alpha}=
F'\left(\mathbf{\mathbf{n}}'\cdot\mathbf{S}\right)\epsilon_{\alpha\beta\gamma}n'_{\beta} S_{\gamma}. \label{derivative-S}
\end{equation}

Comparing  Eqs.\ (\ref{LS-final}) and (\ref{derivative-S}) we immediately see that the time derivative of the total angular momentum, ${\bf J} = \hbar{\bf S} + {\bf L}^{(S)}$, is zero. This concludes derivation of the exact conservation of the total angular momentum, spins + lattice, in the full nonlinear elastic theory with embedded spins. In that derivation we did not use the explicit form of the function $F\left(\mathbf{n\cdot S}\right)$ representing the spin Hamiltonian. The same derivation will apply to any form of the spin-lattice interaction. 

One observation that follows from our derivation is that conservation of the total angular momentum in spin-lattice dynamics can only be recovered after one accounts for all nonlinear terms in the elastic and magnetoelastic parts of the Hamiltonian. This must have consequences for quantum theory of spin-lattice interactions as well. In quantum mechanics deformations are quantized according to
\begin{equation}\label{u-quantized}
\hat{\bf u}({\bf r}) = \sqrt{\frac{\hbar}{2\rho V}}\sum_{{\bf k}\lambda} \frac{{\bf e}_{{\bf k}\lambda}}{\sqrt{\omega_{{\bf k}\lambda}}}\left(e^{i{\bf k}\cdot {\bf r}}a_{{\bf k}\lambda} + e^{-i{\bf k}\cdot {\bf r}}a^{\dagger}_{{\bf k}\lambda}\right),
\end{equation}
where $\omega_{{\bf k}\lambda}$ and ${\bf e}_{{\bf k}\lambda}$ are frequencies and polarizations of phonons, $V$ is the volume of the body, and $a^{\dagger}_{{\bf k}\lambda}, a_{{\bf k}\lambda}$ are phonon operators of creation and annihilation. 

In the linear elastic theory the strain tensor $u_{\alpha\beta} = \frac{1}{2}(e_{\alpha\beta} + e_{\beta\alpha})$ is linear on these operators. The elastic Hamiltonian (\ref{H}) is quadratic on phonon operators while the spin-lattice Hamiltonian in the rotational approximation of Eq.\  (\ref{HS-linear}) is linear on phonon operators. Consequently, at low temperature, when thermal phonons are absent, the spin-lattice theory that describes absorption and emission of phonons is similar to the theory of atomic transitions in electrodynamics. 

This changes when one accounts for nonlinear (anharmonic) terms needed to conserve the angular momentum. The exact strain tensor (\ref{strain-nonlinear}) contains terms that are quadratic on deformation and thus quadratic on the phonon operators. This contributes terms up to the fourth order on phonon operators to the elastic Hamiltonian (\ref{H}). As to the exact spin-lattice Hamiltonian, according to Eqs. (\ref{Exact-Ham_S}) and (\ref{npr_alpha_res}) it contains all orders of the deformation and, thus, all orders of the phonon operators. 

Consequently, multiphonon processes and interaction between phonons, required by the conservation of the total angular momentum, inevitably enter the quantum problem. The intrinsic nonlinearity of the elastic theory makes the dynamics and manipulation of an atomic spin in a solid fundamentally different from the dynamics and manipulation of the atomic states in vacuum. The latter are described by the linear electrodynamic theory that does not require multiphoton processes for the conservation of the angular momentum. 
 
This does not mean that a revision of the theory of spin-phonon transition rates is needed at low temperature. When thermal phonons are absent, contribution of multiphonon processes to the rates would be proportional to higher powers of the spin-lattice interaction and, therefore, would be small. High nonlinearity of this many-body problem, however, makes it difficult to establish with certainty how strongly the phase of the wave function and decoherence of quantum spin states are affected by their entanglement with the infinite number of phonon states mandated by the conservation of the angular momentum in a macroscopic body. 

The prospect of using spins in a 3D solid as qubits \cite{Awschalom} may depend on the answer to this question. The reduction of the number of mechanical degrees of freedom in a nanocantilever-like setup \cite{walmorkab06apl,ganklyrub13NatNano,Freeman-2018,Freeman-PRB2020} may be beneficial in that respect. Similar argument applies to qubits based upon nano-SQUIDs \cite{SQUID}, where angular momentum is generated by the superconducting current \cite{EC-AK-PRB03}. 

This work has been supported by the Grant No. DE-FG02- 93ER45487 funded by the U.S. Department of Energy, Office of Science.

\end{document}